# Realizing topological edge states in graphene-like elastic metamaterials

Zhen Huang[1,2], Penglin Gao[3], Federico N. Ramirez[4], Jorge García-Tiscar[4], Alberto Broach[4], Jiu Hui Wu[2*], Fuyin Ma[2*], José Sánchez-Dehesa[1*]

[1]Wave Phenomena Group, Universitat Politècnica de València, Camino de vera s.n. (Building 7F), ES-46022 Valencia, Spain

[2]School of Mechanical Engineering, Xi'an Jiaotong University, Xi'an 710049, China.

[3]School of Mechanical Engineering and State Key Laboratory of Mechanical System and Vibration, Shanghai Jiao Tong University, Shanghai 200240, China.

[4]CMT Motores térmicos, Universitat Politècnica de València, Camino de vera s.n. (Building 6D), ES-46022 Valencia, Spain.

*ejhwu@xjtu.edu.cn, xjmafuyin@xjtu.edu.cn, jsdehesa@upv.es

The study of topological states in electronic structures, which allows robust transport properties against impurities and defects, has been recently extended to the realm of elasticity. This work shows that nontrivial topological flexural edge states located on the free boundary of the elastic graphene-like metamaterial can be realized without breaking the time reversal, mirror, or inversion symmetry of the system. Numerical calculations and experimental studies demonstrate the robust transport of flexural waves along the boundaries of the designed structure. The topological edge states on the free boundary are not limited by the size of the finite structure, which can reduce the scale of the topological state system. In addition, unlike the edge states localized on the free boundary in graphene where the group velocity is zero, the edge states on the elastic metamaterial plate have propagation states with non-zero group velocity. And there is a frequency range for the edge states, and we introduce the concept of Shannon entropy for elastic waves and use it to assess the frequency range of the edge states in graphene-like elastic metamaterials. This work represents a relevant advance in the study of elastic wave topological states, providing a theoretical basis for engineering applications such as vibration reduction and vibration isolation of mechanical structures.

## I. Introduction

Mechanical vibrations are ubiquitous in many circumstances related to our daily life, such as high-speed trains, airplanes, subways, and other means of transportation. At low frequencies, the excited elastic waves usually produce non-negligible damage to the embedded mechanical devices. Therefore, the design of novel structures for the control of elastic wave propagation acquires paramount relevance. In this aspect, elastic metamaterials are new artificial structures showing exciting properties, including wave transformation[1,2], negative refraction[3,4], focusing, and elastic cloaks[5,6]. As a consequence, the research on elastic metamaterials has been rapidly developed in the last few years.

Topological states were initially studied in condensed matter physics since they exhibit novel properties, such as topologically protected one-way energy propagation without loss and backscattering[7-14]. This property aroused the interest of researchers in extending topological states to the realm of elasticity, combined with the specific features of elastic metamaterials. Thus, the study of topological materials for elastic waves has emerged as a novel research topic in elasticity.

In the field of electronic structures, research on topological states mainly focuses on the quantum Hall effect (QHE)[7,13], the quantum spin Hall effect (QSHE)[9,14-17], and the quantum valley Hall effect (QVHE)[10,11]. Among them, the key to observing the QHE is to break the time-reversal symmetry[7,13], In the field of elastic structures, in order to break the time-reversal symmetry, the elastic analog of the QHE employs moving elements added to the system, such as introducing a rotating gyroscope at each site of the lattice[18,19]. However, the effects brought about by active control systems, such as instabilities and noise, are unavoidable and hinder potential engineering applications. As a result, passive metamaterials are gradually being introduced into the study of topological states[20]. For example, both topological insulators with pseudospin-orbit coupling and valley topological insulators with pseudospin state, rely on the passive properties of metamaterials without breaking the time-reversal symmetry. For the former, spin-orbit coupling is the key ingredient to generate a pair of conjugated topological edge states with opposite spins and protected by time-reversal symmetry[21-28]. As for the QVHE, which was derived from single-valley physics in the realm of elastic waves[25], an effective strategy to open a topologically nontrivial band gap is using point group symmetry conditions[26-32]. It should be noted that the realization of topological states in metamaterials commonly relies on domain walls, which serve as the essential framework for the existence of topological states[33].

In short, the realization of topological states in elastic wave systems requires active or passive methods to change the time-reversal symmetry or the lattice symmetry of crystals to obtain the nontrivial topological phases, which can be quantified by topological invariants such as Chern numbers or (valley) spin Chern numbers. And the edge states are also considered to only appear at the domain wall between the topologically trivial and nontrivial insulators. In addition, previous research on topological states of elastic waves focused on realizing topological band gaps, such as reducing the symmetry of the system[30], changing the distribution of the eigenmode field of the system[25], or producing an effective synthetic gauge flux by structure[34]. In fact, topological phase transitions can also be generated with the bandgap closed, such as the graphene structure in an electronic system, a semi-infinite and a gapless graphene sheet with a zigzag edge has a band of zero-energy states localized at the edge[35-37]. The elastic analog was theoretically proposed in 2013 using a simple model consisting of a honeycomb distribution of mass-spring resonators on top of a thin plate[36], its realization in an actual elastic plate would be of great interest to prove the transmission through edge states in a gapless structure.

Here, we demonstrate the feasibility of topologically protected edge states in a graphene-like elastic metamaterial, without breaking the time reversal, mirror, or inversion symmetry of the system. Edge states appear at the free borders of the system rather than on domain walls formed by metamaterials with different topological phases, producing a significant reduction in the system size and holding great potential for device application. At the frequency of the Dirac point in structures formed by zigzag boundaries, we find frequency bands composed of edge states. Then, by analogy to the electronic propagation in graphene, we confirm that the edge states on the bands have non-trivial topological phases by mapping it to a one-dimensional lattice with chiral symmetry. Numerical calculations and experimental data show that there is a range of frequencies where edge states can be excited in the graphene-like elastic metamaterial ribbon. In addition, in order to measure the frequency range where edge states occur in graphene-like elastic plate, we introduce the concept of elastic Shannon entropy. To the best of our knowledge, the phenomenon of topologically protected flexural wave transport in graphene-like elastic plate has yet not seen the light of day.

## II. Topological states in the graphene-like metamaterial plate

**A. Topological states on the boundary**

To construct acoustic or elastic topological states in phononic crystals, typically two conditions are required, as shown in Fig. 1a. Firstly, the field intensity should exhibit a vortex state, such as vortex pseudospins in the acoustic valley Hall insulators and pseudospin-up or pseudospin-down circulating in acoustic topological insulators[38,39], and secondly, a domain wall needs to be constructed in the phononic crystal[33], where the intensity vortex on either side of the domain wall is in opposite direction to get the topological edge states[38]. Acoustic systems that satisfy these two conditions, such as acoustic valley topological insulators and acoustic topological insulators, exhibit topological-protected unidirectional propagation states on the domain wall. The two sonic crystals that form the domain wall have a bulk bandgap where the edge states are embedded, as shown in Fig. 1a. However, domain-wall systems have usually large dimensions, hindering their integration in engineering applications. Additionally, the edge states can only exist on the interface rather than free boundary. It can be imagined that when two phononic crystals are folded along the domain wall[26], the topological states will appear on the free boundaries of the folded system, as shown in the upper panel of Fig. 1b, but the system becomes more complex. Therefore, further research should be performed with new proposals allowing topologically protected edge states on the free boundaries of simple elastic systems, as shown in the lower panel of Fig. 1b, to simultaneously reduce the size of the structure for miniaturization purposes and improve its integration.

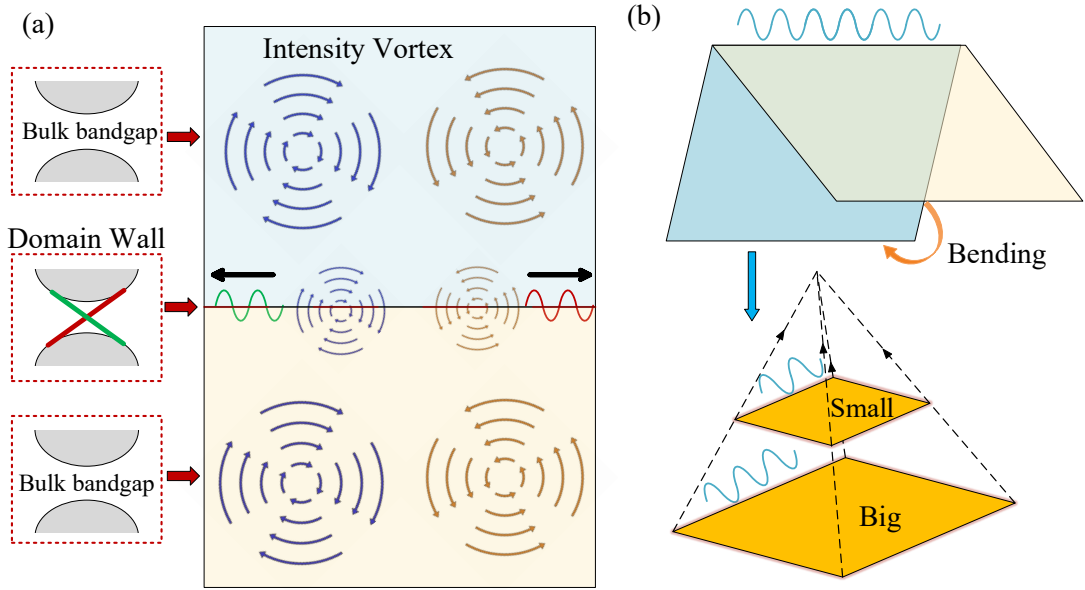

FIG. 1 (a) Schematic pictures representing a topological non-trivial system with domain walls. The phononic crystals on both sides of the domain wall contain a bulk band gap with the same or similar frequency range but different topological phases. This bandgap supports topological edge states localized on the domain wall. (b) Sketches illustrating the topological state propagation in a folded system of elastic crystals along the domain wall and the case on the boundary of a simplify system. The upper panel represents the folded system, while the lower panel represents the simple system, where the dashed arrow indicates the direction in which the big simple system can be reduced into a small one.

## B. Elastic analog of graphene

The electronic properties of graphene, such as the integer QHE and the lossless supercurrent, have been attracting intense interest. These effects are a consequence of the electronic band structure of graphene, which consists of two Dirac cones at the K and K´ points in the *k* space, as shown in Fig. 2. A graphite sheet behaves like a gapless semiconductor due to the electronic modes, which are called edge states, localized near the zigzag edge of graphene[37,40]. These localized electronic states correspond to non-bonding molecular orbitals[41]. The edge states could also exist near the zigzag end of a single-wall carbon nanotube since a carbon nanotube can be considered a graphene sheet wrapped into a cylinder[40]. Researchers explored the zero modes of the graphene tube and found that two perfect mid-gap edge states exist, in which the particle is completely located at the boundary, even for a tube with finite length[42,43]. The honeycomb lattice, defined by colored circles in Fig. 2, can be formed by the repetition of the supercell (gray shaded region) due to the translational symmetry. The supercell of the tube can be reduced to a 1D SSH model[42].

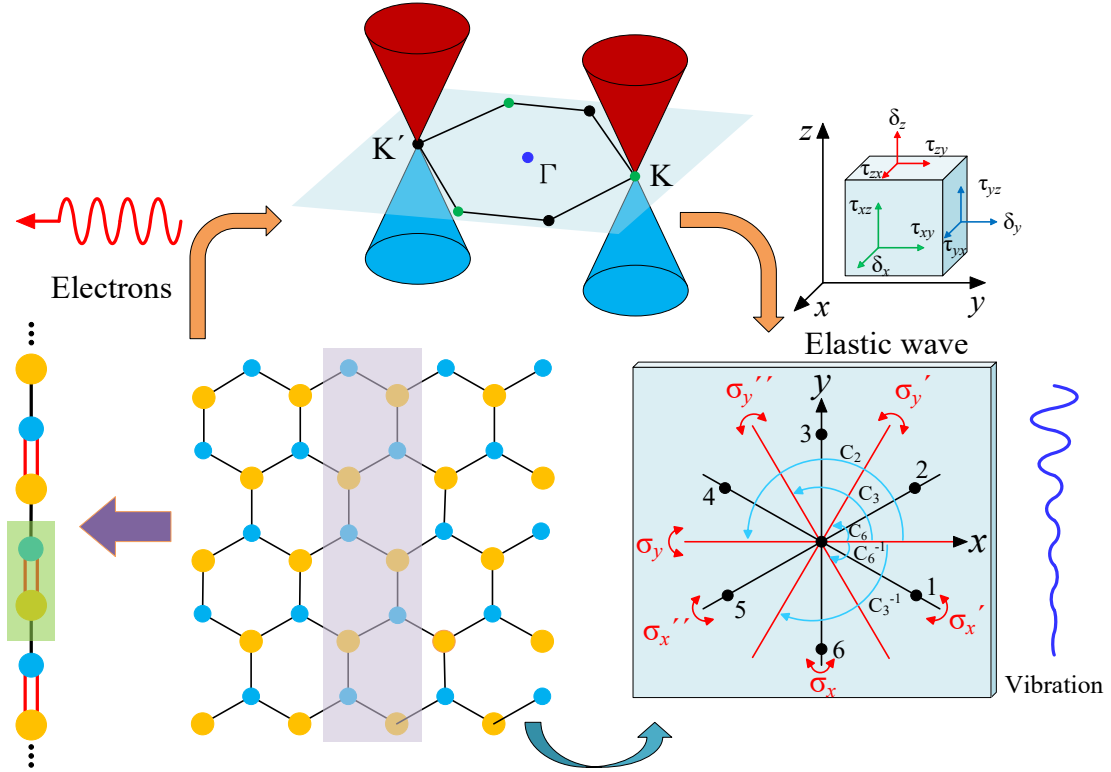

FIG 2. Schematic diagram showing the analogy of graphene edge states to elastic systems

The SSH model, proposed by Su, Schrieffer, and Heeger[44], is the prototype of a topologically nontrivial band insulator with a symmetry protected topological phase. Therefore, we can extend the mid-gap edge states in electrons to elastic waves in solids based on the Dirac cones at high symmetry points K and K´ in the Brillouin zone. Dirac cones are characterized by two principle features: the double degeneracy and linear dispersion near the degenerate point, both depend on the symmetry of the crystal lattice[45,46]. However, because of the full-vector feature, and the complicated couplings between the longitudinal and transverse components of elastic waves, manipulating elastic waves in solids, such as rapidly decaying wave amplitude from a solid boundary to the interior, is generally challenging[26]. Our work mainly considers the edge states of flexural waves in thin plates with triangular lattice with symmetry operations of the $C_{6v}$ group, as shown in Fig. 2. In this case, we can concentrate the vibration of the elastic plate on the edge, and the vibrational energy is rapidly attenuated with the distance to the edge, as it is depicted schematically by the blue curve in Fig. 2.

**C. Band structure of the unit cell**

Let us consider an elastic plate patterned with a periodic array of hexagonal holes containing three one-beam resonators equally distributed inside the hexagonal cavity, as shown in Fig. 3a. We select this unit cell since their parameters can be adjusted to obtain specific properties of the flexural band structure[26]. The upper and lower panels show top and cross-sectional views of the unit cell, respectively. The lattice constant of the cell is $a$. Other geometric parameters include the plate thickness $h$, radius $R$ of the inscribed circle in the hexagonal hole, diameter $d$ of the cylindrical pillar, cylinder height $t$ outside the plate, width $w$ of the beam, and the center moment $\ell$ of the beam.

In order to study the effects of the one-beam resonators on the band structure, we first analyze the dispersion relation of the unit cell without resonators, which is shown in Fig. 3b. The hexagon in Fig. 3b represents the first Brillouin zone (BZ), whereas the orange region M-K-Γ-M represents the irreducible BZ. The bottom inset depicts a scheme of the unit cell. This work mainly analyzes the propagation behavior of the flexural wave in the elastic plate. The governing equation for the displacement of the triangular lattice plate is[26]: $(\lambda + \mu)\nabla(\nabla \cdot \boldsymbol{U}) + \mu\nabla^2 \boldsymbol{U} = \rho\ddot{\boldsymbol{U}}$, where $\lambda$ and $\mu$ are Lame constants, and $\boldsymbol{U}(x,y,z) = u\boldsymbol{i} + v\boldsymbol{j} + w\boldsymbol{k}$ denotes the displacement vector; $\rho$ is the material density. Therefore, we define the out-of-plane polarization ratio $D_Z$ to distinguish the flexural modes, which can be calculated by[47]: $D_Z = \frac{\int_V z^2 dV}{\int_V (x^2+y^2+z^2)dV}$, where, $x$, $y$, $z$ are the components of displacement along the $x$, $y$, and $z$ directions, respectively. The symbol $V$ represents the volume of the unit cell. The parameter $D_Z$ takes values from zero to 1. When the value $D_Z$ is close to 1, it means that the vibration of the elastic plate is a pure flexural wave. However, when $D_Z$ is close to zero, it means that the vibrations are in-plane. For further details of the numerical simulation and the convergence of the calculated frequencies, see Sec. I in **Supplement Material**. Blue and red dotted lines represent the out-of-plane and in-plane modes, respectively, in the band structures shown in Fig. 3. Figure 3b shows that a deterministic Dirac cone arises at the corner K in the BZ as a result of the $C_{6v}$ symmetry of the unit cell, consisting of sixfold rotational and six inversion symmetry in the *x-y* plane. And there is another Dirac point at K' derived directly from time-reversal symmetry[38]. However, the Dirac point P is not at the valley of the two related bands, so an additional structure should be proposed to accomplish such condition.

In order to adjust the band structure, we use the known properties of local resonance excited by a lattice of one-beam resonators[48]. Figure 3c represents the band structure calculated with the unit cell shows in the inset. It is observed that the Dirac point P at the band valley still exists, but it deviates from the corner of the BZ because of the reduced lattice symmetry. It is worth noting that compared with Fig. 3b, the band structure in Fig. 3c shows three additional bands caused by the lattice of one-beam resonators, and all of them contain frequencies where the dispersion relations are flat due to the lattice of local resonances (see Sec. II in **Supplement Material**). To study the characteristics of these three extra bands, we have studied the band eigenmodes at the positions named A, B, and C in Fig. 3c; all located at the corner K of the BZ. Figure 3d plots the eigenmodes, showing that they correspond to fundamental modes of the one-beam resonators, i.e., flexural mode (A), translational mode (B), and torsional mode (C). In addition, Fig. 3d also depicts the spin mode D, which is located at a much higher frequency (not shown in the band structure). To decrease the frequency position of the Dirac point, we realize that the flexural band structure could be adjusted by just tuning the parameters defining the one-beam resonators. For example, for the band gap (gray area) in Fig. 3c, a simple mass-spring-pendulum model can be used to explain its formation mechanism, i.e., due to the introduction of the pendulum, the resulting band structure of the 1D system formed by the combination of mass and spring has two branches. Therefore, a gap appears between the maximum frequency of the first branch at the zone boundary and the lowest frequency of the second branch at the Γ point[49]. Furthermore, the central frequency of the resonance bandgap can be obtained by modeling the one-beam resonator to a simple spring-mass structure (as shown in the inset in Fig. 3e), and the expression is $f_M = \frac{1}{2\pi}\sqrt{\frac{K_e}{M_e}}$, where $M_e$ and $K_e$ represent the equivalent mass of the spring-mass oscillator and the equivalent stiffness of the spring, respectively. Since the beam is modeled as a spring, the equivalent mass is the mass of the cylinder at the end of the beam $M_e = \pi r^2 (t + h)\rho$, where the radius of the pillar $r = d/2$. Regarding the stiffness of the equivalent spring is $K_e = EI(\frac{1}{2}L_0 L^2 - \frac{1}{6}L^3 + (L_0 - L)(L_0 L - \frac{1}{2}L^2))^{-1}$, where *I* represents the moment of inertia of the rectangular cross-section beam $I = wh^3/12$; *L* is the effective length of the beam $L = l - r$, and $L_0$ represents the equivalent moment of the concentrated force action point $L_0 \in (l, l+r)$. Thus, the central frequency, $f_M$, of the bandgap associated with one-beam local resonance can be approximated by the simple model.

To guarantee the robustness of the structure to be manufactured, we fix the width of the beam to the value $w$=2mm. Therefore, we study the dependence of $f_M$ with the length $L$ of the beam, the radius $r$ of the disk, and the height $t$ of the cylinder added on top of the disk. The orange curve in Fig. 3e indicates the dependence with the cylinder radius when $L$=4mm and $t$=1mm. The green curve depicts the dependence with $L$ when $r$=3mm and $t$=1mm, and the blue curve represents the relationship between the extra height $t$ of the cylinder and the intermediate frequency of the band gap when $r$=3mm and $L$=4mm. It is observed that $f_M$ decreases with the increase of $L$, $r$ and $t$. From the slopes of the three curves, it can be seen that the radius of the disk has the greatest influence on $f_M$ followed by the length of the beam, and finally, the height of the cylinder added to the disk.

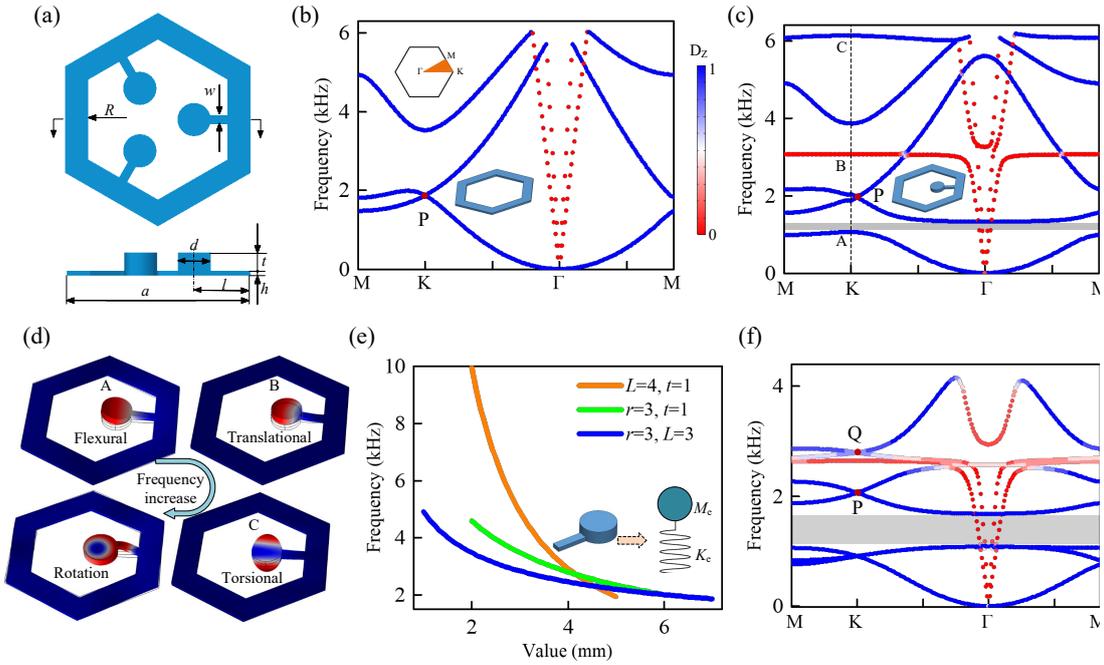

FIG. 3 (a) Schematic diagram of the unit cell with the definition of its geometrical parameters. The upper figure shows a top view, and the lower figure is a cross-sectional view. (b) Band structure of a unit cell without the three-beam resonators. The upper inset depicts the first Brillouin zone, and the lower represents the cell. The blue and red dotted lines indicate the out-of-plane and the in-plane modes, respectively. (c) Band structure of a hexagonal lattice where the unit cell contains a one-beam resonator as described in the inset. (d) Displacement field distributions for the eigenmodes of local resonances A, B, and C in (c). The blue arrow indicates the modes with increasing frequency. (e) The behavior of the mid-frequency resonant bandgap produced by the one-beam resonator embedded in the unit cell. The inset represents the one-beam resonator, which is made equivalent to a spring-mass system. (f) Band structure corresponding to the unit cell shown in (a).

To ensure that a deterministic Dirac point appears at point K, we employ the unit cell shown in Fig. 3a, where three one-beam resonators are symmetrically distributed in the unit cell, so that the lattice and the high symmetry point K have $C_{3v}$ symmetry at the same time[45]. In this case, where three one-beam resonators are considered, the number of resonant modes is triple in comparison to the case of lattices with only one-beam resonator. To have the central frequency of the bandgap, $f_M$, smaller than the frequency of the Dirac point P in Fig. 3b, we choose a set of parameters extracted from Fig. 3e (**Methods**). Figure 3f shows the calculated band structure, where the Dirac point P, with frequency 2068.7 Hz, occurs at the high symmetry point K. The local modes introduced by the three one-beam resonators produce three flexural flat bands at low frequencies. A bandgap occurs between the lower two flat bands that are degenerated at Γ-point and the third one. The low-frequency bandgap caused by local resonance can achieve the goal of low-frequency vibration reduction, so the thin plate considered here can be called an elastic metamaterial[5]. It is worth noting that there is another Dirac point Q at corner K. In comparison with the band structure associated with the one resonator case (see Fig. 3c), the additional Dirac point Q is caused by the in-plane (translational) resonant mode arising from the three one-beam resonators. The underlying physics involves the interaction of the in-plane mode and the flexural waves propagating in the plate, which produces a fully out-of-plane mode, defining a Dirac cone under the protection of structural symmetry.

**D. Dispersion relation of ribbons**

To analyze the edge states induced by Dirac cones, we realize that they are two types of interfaces in an infinite elastic metamaterial slab. Figure 4a shows a scheme of the sample containing the two kinds of interfaces; the armchair (AM) interface, and the zigzag (ZZ) interface. First, let us study finite "ribbons" formed by the AM interfaces as shown in the supercell depicted in Fig. 4b. The ribbon is considered periodic along the *y*-axis, with lattice constant $A=\sqrt{3}a$, and contains 7 cells of type I (see inset of Fig. 4b) in the direction defined by the *x*-axis. Figure 4c shows the calculated dispersion relation of the flexural modes (blue dotted lines) where an omnidirectional bandgap of out-of-plane modes (gray stripe) appears around the frequency of the Dirac point. Since flexural waves with frequencies in the bandgap cannot propagate in the ribbon, the calculated flexural modes correspond to states propagating in the bulk of the ribbon, and no topological edge states appear in this ribbon. Let us point out that in-plane modes (red dotted lines) can also propagate in the ribbon but they are of no interest to this work.

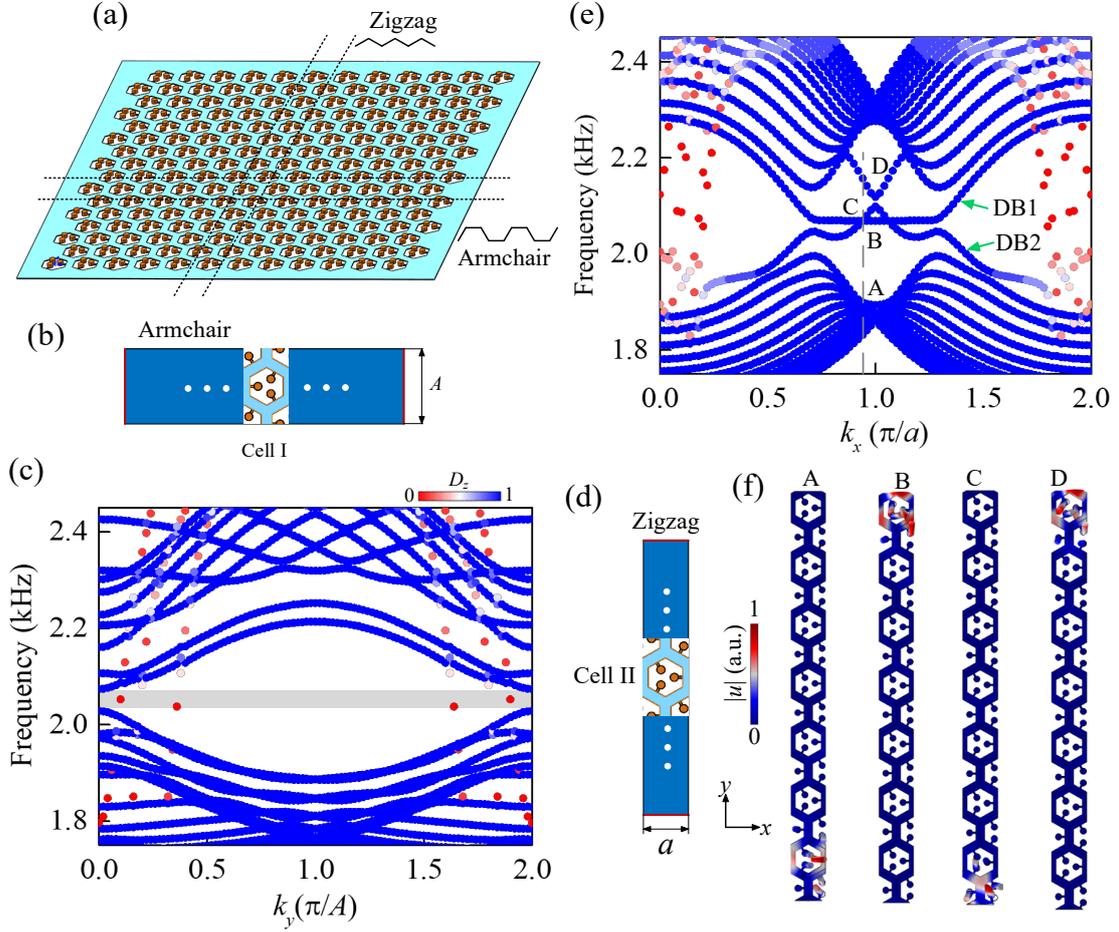

FIG. 4. (a) Plot of the designed metamaterial plate employed to study flexural edge states. (b) Ribbon containing armchair boundaries. (c) Dispersion relation of the projected bands corresponding to the ribbon described in (b). Green in the band structure indicates out-of-plane modes, while red indicates in-plane modes. (d) Ribbon containing zigzag boundaries. (e) The dispersion relation for the ribbon (d). (f) Distribution of displacement field for the four eigenmodes at $k_x$=0.94, A, B, C and D in (e). Color represents total displacement $|u|$.

Next, we analyze ribbons with ZZ edges, which are obtained with the supercell shown in Fig. 4d. The ribbon is periodic along the *x*-axis, with lattice period *a*, and contains 7 cells of type II (see the inset) in the *y*-axis direction. Figure 4e shows the corresponding dispersion relation for the flexural modes (blue dotted lines). It is observed that, in addition to the bulk modes, there is a pair of bands (DB1 and DB2) extending near the Dirac point, similar to the zero-energy modes in the graphene energy band structure[50]. However, unlike graphene sheets with two flat bands near $k_x$=1[37], or the elastic granular graphene without band of edge modes due to the breaking of chiral symmetry on the zigzag interface[51], the band DB2 has a non-zero group velocity near $k_x$=1, so the two bands, DB1 and DB2, cross in $k$ space. Moreover, two isolated bands also appear on the upper and lower sides of the two edge-state bands near $k_x$=1, due to the free boundary. We select four eigenmodes at $k_x$=0.94 near the frequency of Dirac point, A, B, C and D, as shown in Fig. 4f, the total displacement $|u|$ of the four modes exhibit a strong localization at top or bottom end of the ribbon, decaying rapidly into the bulk of the ribbon. These localized mode has the features of a flexural wave[52]. Specifically, modes A and D, are boundary states caused by free boundary conditions, while modes B and C are singular edge states analogous to graphene[35,37], in addition, the characteristic of the mode C on DB1 is that it is localized at the bottom end of the ribbon, while the mode B on DB2 is localized at the top end of the ribbon.

**E. One-dimensional model and topological phase**

To check that edge states have topological protection, the assessment method consists of calculating the topological phase in the corresponding frequency band. Typically, nontrivial topological phases of 2D periodic structures are embodied in the bandgap, and topologically protected states appear at the interface formed by phononic crystals of different topological phases[25]. Therefore, it is challenging to calculate the topological phases of gapless structures, due to the band degeneracy (Dirac cone) as in the case studied here. We adopt the winding number[43], which is an integer representing the total number of times that a closed curve travels counterclockwise around a given point so that the wave function remains unaltered, to discuss the topological arguments[52]. Figure 5a demonstrates a 2D infinite metamaterial plate, where $a_1$ and $a_2$ are the primitive lattice vectors of the sheet, which determine the shape of the free boundary of the elastic plate[53]. The yellow dashed box in Fig. 5a defines the reduced 1D ribbon, which recovers the entire 2D metamaterial plate by translations along the $a_1$ direction. The smallest unit cell of the reduced 1D crystal, Cell III, is the honeycomb cell indicated by the green diamond area in Fig. 5a, where $b_1 = (1/2, \sqrt{3}/2)a$ and $b_2 = (1/2, -\sqrt{3}/2)a$ are the corresponding lattice vectors. Therefore, the vectors of the reciprocal lattice are $d_1 = 2\pi/a\,(1, -1/\sqrt{3})$ and $d_2 = 2\pi/a\,(1, 1/\sqrt{3})$, as shown in Fig. 5b.

For the honeycomb lattice, the Hamiltonian can be calculated based on the tight-binding model[50,54], which is $H(\boldsymbol{k}) = [0, g(\boldsymbol{k});\ g^*(\boldsymbol{k}), 0]$. By considering only the relative phase to adjoining sites, the off-diagonal term of the Hamiltonian can be expressed as $g(\boldsymbol{k}) = 1 + e^{i\boldsymbol{k}\cdot(\boldsymbol{\delta}_3-\boldsymbol{\delta}_2)} + e^{i\boldsymbol{k}\cdot(\boldsymbol{\delta}_3-\boldsymbol{\delta}_1)}$, in which, $\boldsymbol{\delta}_i$, $i$=1, 2, 3 are the nearest-neighbor vectors[53]. The Bloch wave vector $\boldsymbol{k}$ should be expressed in accordance to the honeycomb lattice $\boldsymbol{k} = k_h \boldsymbol{d}_1 + k_v \boldsymbol{d}_2$, $k_{1,2} \in [0,1)$. Substituting the expression $\boldsymbol{k}$ into $g(\boldsymbol{k})$ finally gives rise to $g(k_h, k_v)$. The two-by-two effective Hamiltonian $H(\boldsymbol{k})$ can be used to describe a 1D lattice model when $k_v$ runs through the 1D BZ for each $k_h$. Thus, the winding number can be characterized by $e(k_h) = \frac{1}{2\pi i}\int_0^1 \frac{\partial \ln[g(k_h,k_v)]}{\partial k_v} dk_v$, where $g(k_h, k_v)$ graces out a closed curve on the complex plane, and $v(k_c)$ counts the number of times it winds around the origin.

In addition, the chiral symmetry also ensures a quantized Zak phase[53] $\theta(k_h) = \int_0^1 i\langle\psi(k_v)|\nabla_{k_v}\psi(k_v)\rangle dk_v$, which relates to the winding number through $\theta(k_h) = |v(k_h)|\pi$. The blue line (Theory) and red dot (FEM) represent the Zak phase obtained from theoretical derivation and simulation calculation (see sec. III of **Supplementary Materials** for details), respectively, as shown in Fig. 5c, which indicates a nontrivial topological phase for the ZZ edge. Particularly, $\theta(k_h) = \pi$ accounts for a topologically nontrivial phase for 1/3 < $k_h$ < 2/3. The degenerated bands cover one-third of the Brillouin zone. The borders correspond to the transition points when the 1D BZ passes through the reciprocal lattice sites K and K′. It is observed that the edge states between the unequal corners K and K′ have topological protection properties.

The existence of edge states is not affected by the length of the finite ribbon, and there are highly localized edge states even in the ribbon composed of two cells of Cell III (see Sec. IV in **Supplementary Materials** for details). To demonstrate the propagation features of the edge states in a 2D ZZ ribbon, a simulation is conducted with the metamaterial plate described in Fig. 5d, which consists of five cells (type III) in the ribbon, and twelve ribbons along the *x*-axis. For the sake of easier manufacturing of this structure, the one-beam resonators on the left and right sides of the metamaterial boundaries are not considered in the calculations since their contribution is negligible. We use a punctual source located at the position defined by the green star to excite a harmonic flexural wave with frequency *f*=2080Hz. As shown in Fig 5c, the excited wave propagates along the edge of the plate and decays almost exponentially into the interior. The color in the figure represents the distribution of normalized velocity amplitude. According to the previous discussion, edge states exist not only at the frequency defined by the Dirac point but also in a range of frequencies around the Dirac point.

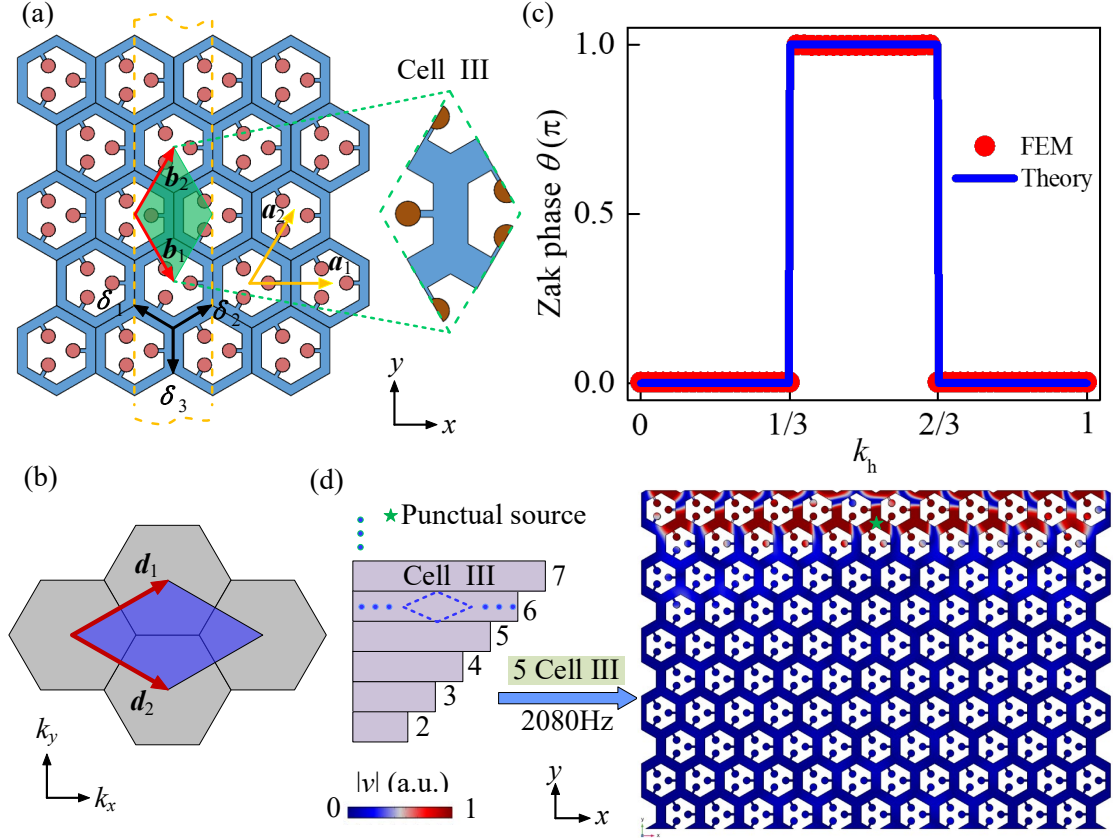

FIG. 5 (a) Two-dimensional elastic metamaterial plate with triangular lattice. The yellow dashed box defines the reduced one-dimensional crystal cell, while the green area indicates the unit cell of the reduced one-dimensional crystal, which is denominated as Cell III. (b) The reciprocal space of the honeycomb lattice. (c) Zak phase of the 1D model obtained from theoretical calculations (blue line, Theory) and simulation calculations (red dot, FEM). (d) Different number N of cells III in the ribbon supporting edge states and the propagation in the plate by a punctual source with a frequency of 2080 Hz. The color represents the distribution of normalized velocity amplitude and the green star indicates the location of the punctual source.

## F. Assessment of edge states by Shannon entropy

Entropy, which is the (log of the) number of microstates or microscopic configurations, so far, had been a concept in physics. In simple terms, if the particles inside a system have many possible positions to move around, the system has high entropy, and if they remain motionless, then the entropy of the system is low. Therefore, from the perspective of spatial distribution, the local and non-local characteristics of eigenmode distribution can also be characterized by entropy. Shannon proposed a computational method to quantify informational entropy, also known as Shannon entropy[55], which was first extended to atomic physics as[56] $S_u = -\int \rho_e(\mathbf{r}) \ln \rho_a(\mathbf{r}) d\mathbf{r}$, where $\rho_e(\mathbf{r}) = |\psi(\mathbf{r})|^2$ is the probability density distribution of a given electronic orbital and more recently to acoustics[57]. Following the proposal in acoustic[57], we introduce here the "probability density function" for our metamaterial elastic plate: $P_s(\mathbf{r}) = V|u(\mathbf{r})|^2 / \int |u(\mathbf{r})|^2 d\mathbf{r}$, where $|u(\mathbf{r})|^2$ is the square norm of the total displacement field, which can be calculated by the dot product of the total displacement and its complex conjugate; $V$ is the volume of the integration domain. The square norms of displacement are proportional to respective field intensities. Therefore, $P_s(\mathbf{r})$ plays the same role as $\rho_e(\mathbf{r})$ in electronic states. Following the definition of Shannon entropy in acoustics[57], to assess the spreading of localized elastic modes we introduce the Shannon entropy as $S_u = -\int P_s(\mathbf{r}) \ln P_s(\mathbf{r}) d\mathbf{r}$. This quantity provides a measure of the spatial delocalization of the modes in the system. The characteristic of Shannon entropy is that $S_u$ increases with increasing uncertainly (i.e. increasing spreading of the displacement field characterizing the eigenmode)[57].

The topological edge states in the ribbon are localized modes, so we use the elastic Shannon entropy to assess the range of frequencies where edge states can be found. Let us remark that in actual calculations, the volume in Shannon's entropy formula refers not to the total volume of the supercell, but to the volume of a single unit cell, which is proportional to the total volume. Since the two corners K and K' that support the Dirac point are located at the 1/3 and 2/3 segment points of the projected BZ respectively, in order to analyze the characteristics of the frequency band between the two corners K and K', without loss of generality, the larger wavevector interval selected here is from $k_x = 0.5$ to $k_x = 1.5$. For convenience, within the range of $k_x$ from 0.5 to 1.5, the two segment bands are named UM in the band DB2 (with the edge states located at the top end of the ribbon), and DM in the band DB1 (with the edge states located at the bottom end of the ribbon), respectively, as shown in Fig. 6a. It can be seen from the figure that for the DM segment, in the interval between $k_x = 0.7$ and 1.3, which is approximately the 1/3 and 2/3 dividing points of the 1D projection space, the frequency band is a flat band with a zero group velocity similar to that of graphite[37]. However, in the UM segment, the edge states localized at the top of the ribbon has a non-zero group velocity.

The Shannon entropy on the bands UM and DM is shown in Fig. 6b. It is observed that Shannon entropy exhibits a symmetric distribution with respect to $k_x=1$. As Shannon entropy increases, the vibrational modes extend into the supercell. Therefore, near $k_x=1$, the eigenstates are highly localized. In order to determine the frequency range of the edge states, we assume that the displacement field of the edge states is localized within one cell near the end of the ribbon. In this case, when $|k_x - 1| \leq 0.2$, the eigenmode displacement field of the edge states satisfies this assumption, therefore the range of wave vector $k_x$ for the edge states is [0.8, 1.2]. Figure 6c shows the displacement fields of $D_1$ and $T_1$ at $k_x = 0.8$, and $D_2$ and $T_2$ at $k_x = 1$. From the distribution of four models, the vibrational energy is mainly localized in one cell at the upper or lower ends of the ribbon (the ribbon rotates counterclockwise by 90° in the figure). According to the Fig. 6b, $D_1$ and $D_2$ are located in the band with non-zero group velocity, while the group velocity of $T_1$ or $T_2$ is close to zero, indicating that the frequencies of $D_1$ and $D_2$ can be used for signal propagation. Therefore, for the case of vibrational propagation, we consider these edge states with non-zero group velocity. It can be inferred that the frequency range of the edge states lies between $D_1$ and $D_2$, [2034.3Hz, 2097.8Hz], as shown in the shaded area in Fig. 6a. Interestingly, the frequency of the Dirac point is located near the center of the range. It is worth noting that, according to the assumption, the starting point of the edge state is D1, and the Shannon entropy of the displacement field of mode D1 is the same as that with $T_1$. Therefore, it can be concluded that the dashed line, as shown in Fig. 6a, defines the wave vector range of the edge states. Under this criterion, the frequency range of the edge states for the gapless elastic plate can be determined. Additionally, the frequency of the edge states is influenced by the bulk state, so the actual frequency range of the propagating wave along the boundary may slightly differ from the predicted values.

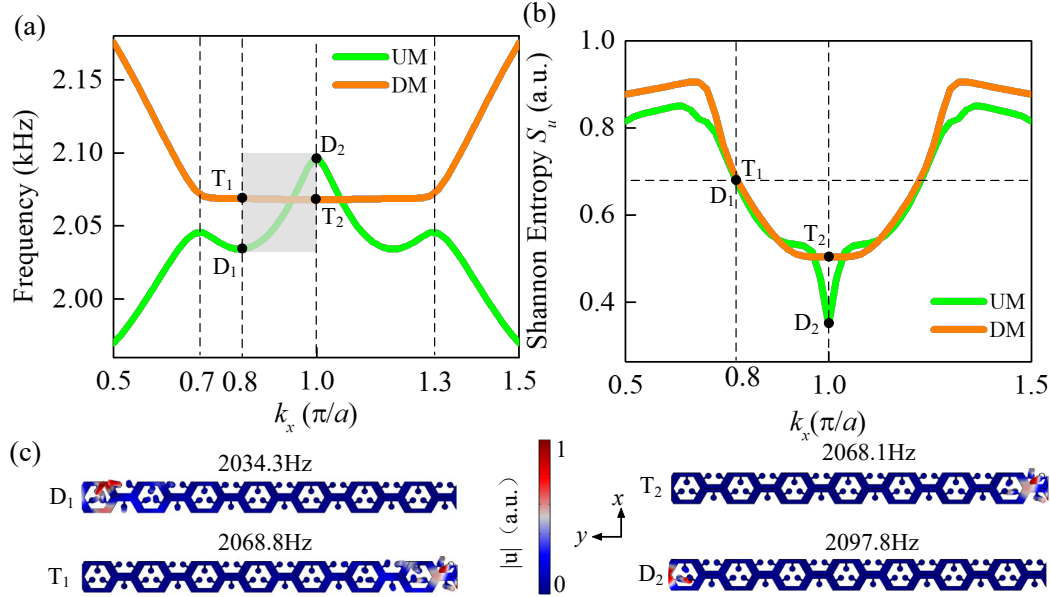

FIG. 6 (a) Shannon entropy of edge modes belonging to the bands UM and DM. (b) Band structure for the bands UM and DM. (c) Displacement fields of eigenmodes $D_1$, $T_1$, $D_2$ and $T_2$ selected in (a).

### III. Experimental observation of edge states

To verify the theoretical predictions regarding the propagation of flexural waves along the boundaries of the metamaterial plate under study we use the experimental setup described in Fig. 7a. In brief, we employ a Polytec PSV-500 scanning laser Doppler vibrometer (LDV) located 0.8 meters away from the metamaterial sample. Figure 7b shows the sample plate, with a dimension of about 480 × 350mm, which has been manufactured by laser-cutting using aluminum (Al-5745) with the same parameters (Young modulus and mass density) as that employed in the simulations. The accuracy of the laser-cutting technique used in manufacturing is about 0.1mm[58]. The prototype was obtained from two aluminum sheets of 1mm and 4mm respectively; the designed metamaterial was cut in the first and cylindrical pillars 4mm thick were obtained from the later. Afterward, the pillars are attached with cyanoacrylate to their positions in the 1mm thick plate. The sample is installed vertically, hanging from two thin wires. In addition, mastic tape was added to the lateral boundaries to avoid possible wave reflections, thus mimicking the absorbing boundary condition employed in the numerical simulations.

Measurements are performed on the flat regions of the metamaterial sample plate (avoiding the cylindrical pillars), which have been discretized in roughly 1250 points. A computer audio card is employed to synthesize a sinusoidal signal, which is amplified with a 200W Samson signal amplifier. The amplified electrical signal feeds a TDK PS1550L40N piezoelectric transducer (PZT), with a diameter of 15 mm and a thickness of 1.6 mm, located as shown in Fig. 7a. The transducer excites flexural waves in the sample plate with the desired frequency. The measurements at each position of the grid were performed at a sampling rate of 16 kHz for 500 milliseconds, repeated three times per sampling point in order to get an averaged result. The LDV system captures the propagation of excited flexural waves within the frequency range from 0 to 6.4 kHz which, using 3200 FFT lines, results in a frequency resolution of 2 Hz. Moreover, a bandpass filter has been applied to the captured signal to filter out any unwanted interfering signal.

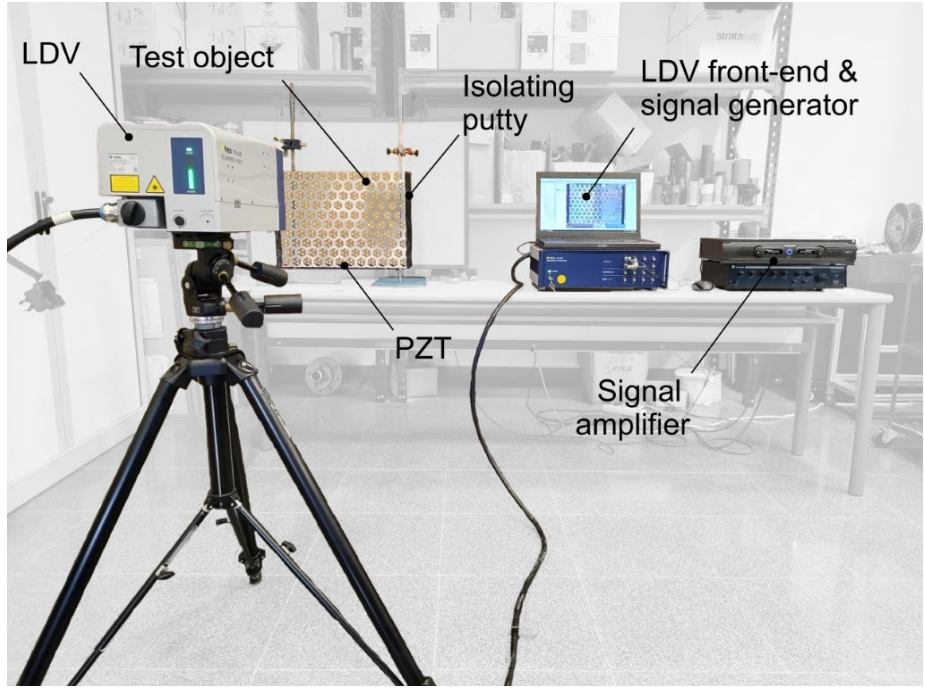

FIG. 7 Experimental setup employed to characterize the flexural edge states propagating in the metamaterial plate under study.

The distribution of out-of-plane velocity fields is depicted in Fig. 8 as the excitation source frequency varies from 1990Hz to 2070Hz in increments of 10Hz. The figure illustrates that when the excitation frequency is at 1990Hz and 2060Hz, the flexural vibration can extend throughout the entire metamaterial plate. However, when the excitation source frequency is approximately in the range of 2010Hz to 2050Hz (with a frequency interval of 40Hz), the flexural waves primarily propagate locally at the free boundary. The experimental data demonstrate fairly well that the topologically protected flexural edge states can propagate along the free boundaries of metamaterial plates[53].

The experimental test results in Fig. 8, when compared to the results obtained from finite element calculations (see Sec. V in **Supplementary Materials** for details), reveal that although the bandwidth of the topological edge states is roughly the same, the experimental results show that the frequency of the edge states shifts down. The discrepancy can be explained by three factors. On the one hand, the added glue increases the mass to the equivalent spring-mass model. On the other hand, slight differences might exist between the aluminum parameters used in the simulations and the ones of the actual aluminum plate. Finally, the number of finite elements used in the simulations just gives an approximation of the converged frequency. In addition, the simulations use low reflection conditions (without mass) at the lateral sides of the plate. However, the experimental setup employs an absorbing material (with mass), which also produces a decrease in the measured frequencies.

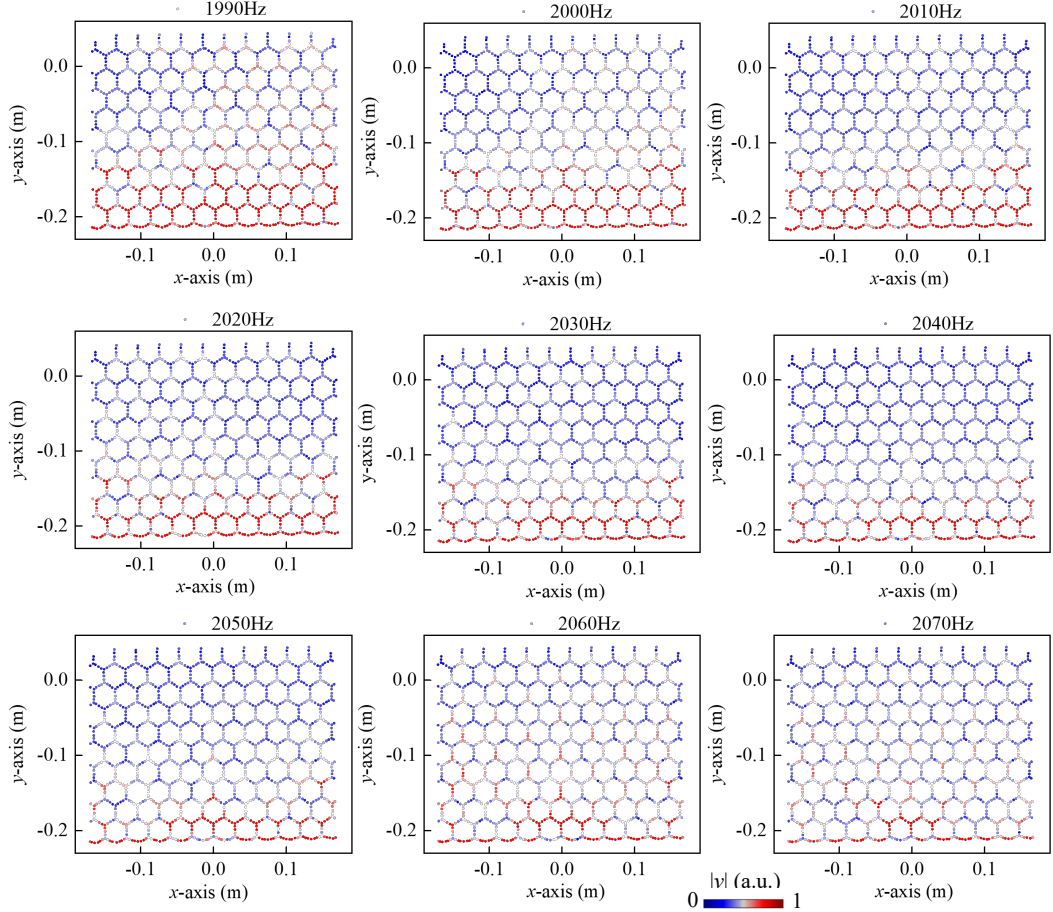

FIG. 8 Snapshot of the out-of-plane velocity field showing the propagation along the boundary of the edge state with frequencies from 1990 Hz to 2070Hz.

## IV. Discussion

In summary, we have demonstrated both theoretically and experimentally the existence of topologically protected edge states at the free boundaries of a graphene-like metamaterial plate, without breaking the time reversal, inversion symmetry of the system. The topological edge states on the free boundary are not limited by the size of the finite structure, which can reduce the scale of the topological state system. And the range of the topological edge states can be assessment by elastic Shannon entropy. Our work can be extended to other elastic waves, such as shear waves, surface waves, etc. The results shown here foresee interesting engineering applications such as non-destructive testing and vibration isolation and open up new avenues for the study of topologically protected edge states in other classical waves such as electromagnetic waves.

# Methods

## Numerical simulations and structure parameters

The numerical results presented in our work are performed in the framework of the fine-element method, using the commercial software, COMSOL Multiphysics. In addition, we employ the "Structural Mechanics module" to visualize the modal characteristics and propagating features of elastic waves. The material of the elastic plate is aluminum with the following physical parameters: Young's modulus $E$=70GPa, Poisson's ratio $v$= 0.33, and mass density $\rho$=2700kg/m$^3$. In the calculations of band structure, Floquet periodic boundary conditions are imposed on the periodic boundaries of the unit cell and the supercell. In the simulation of vibration propagating in the ribbon, we impose low-reflection boundary conditions on the left and right boundaries of the ribbon, and the top and bottom boundaries are free. To calculate the Zak phase, the displacement field of the eigenstates at specific wave vectors is extracted from the numerical simulation of their eigenvalues. The parameters for Fig. 3b are $a$=40mm, R=0.35a, and h=1mm. In Fig. 3c, the parameters defining the resonator are $w$=2mm, $\ell$=10mm, $t$=1 mm, and $d$=8 mm. Fig. 3f is calculated by the following structural parameters: R=0.75, $r$=3.6, and $L$=3.6.

## Data availability

The data which support the figures and other findings within this paper are available from the corresponding authors upon reasonable request.

## Acknowledgements

This work was supported by the National Natural Science Foundation of China (NSFC) under Grant No. 52250287. Zhen Huang acknowledges a scholarship provided by China Scholarship Council (CSC) under Grant No. 202206280162. JS-D acknowledges the support of the RDI grant PID2020-112759GB-I00 funded by MCIN/AEI/10.13039/501100011033. The authors acknowledge Francisco Cervera at UPV for technical help.

Supplementary Material for

# Realizing topological edge states in graphene-like elastic metamaterials

Zhen Huang[1,2], Penglin Gao[3], Federico B. Ramirez[4], Jorge García-Tiscar[4], Alberto Broatch[4], Jiu Hui Wu[2], Fuyin Ma[2], José Sánchez-Dehesa[1]

[1]Wave Phenomena Group, Universitat Politècnica de València, Camino de vera s.n. (Building 7F), ES-46022 Valencia, Spain

[2]School of Mechanical Engineering, Xi'an Jiaotong University, Xi'an 710049, China.

[3]School of Mechanical Engineering and State Key Laboratory of Mechanical System and Vibration, Shanghai Jiao Tong University, Shanghai 200240, China.

[4]CMT Motores térmicos, Universitat Politècnica de València, Camino de vera s.n. (Building 6D), ES-46022 Valencia, Spain.

*ejhwu@xjtu.edu.cn, xjmafuyin@xjtu.edu.cn, jsdehesa@upv.es

## I. Flexural Wave Velocity and Finite Element Mesh

Considering only linearly elastic displacements, away from the source, the propagating waves in an elastic plate are governed by Lamb's homogeneous equation, whose solutions are called Lamb waves. When the wavelength is much larger than the plate thickness, a simpler set of governing equations derived from classical plate theory can be used to understand the motion. For thin plates, the waves have two modes of propagation. One is called the extensional and the other the flexural mode. Both have in-plane and out-of-plane components due to the Poisson effect. For the extensional mode, the larger component of its two displacement components is in-plane, while the larger component in the flexural mode is perpendicular to the plane of the plate[1].

For an isotropic plate, the extensional mode is analogous to the extensional wave in a rod, and just as in the case of the rod, it is dispersionless. The (velocity) dispersion equation is given by

$$c_e = [E/(1-v^2)\rho]^{1/2} \tag{1}$$

Where $E$ is Young's modulus, $v$ is Poisson's ratio, and $\rho$ is the density.

The flexural wave dispersion equation for the plate is analogous to flexural waves in a Euler beam and the velocity shows a similar dependence on the frequency

$$c_f = (D/\rho h)^{1/4} \omega^{1/2} \tag{2}$$

Where $D = Eh^3/12(1-v^2)$, $\omega$ is the circular frequency (rad/s), and $h$ is the plate thickness. Note that the velocity, $c_f$, of flexural (bending) waves is frequency dependent. In addition, since the theoretically obtained bending wave velocity does not consider the inertial effect and the shearing effect, there is a large deviation between the experimentally obtained bending wave velocity and the theoretical value[1].

For the flexural wave modes, the dispersion relation still varies as the square root of the frequency. Here, when referring to waves, the term flexural means displacement out-of-plane.

It is important to point out that the wave speeds are different for the types of modes propagating in the plate. For the in-plane modes, the wave speed in an aluminum plate is about twice as fast as the flexural mode, at the highest flexural frequency measured, and, in the composite laminate, it was about five times as fast as the flexural mode. Combined with the nondispersive nature of the extensional mode, this enabled the modes to be identified quite easily.

Considering the uncertainty of the vibration propagation speed in the plate, especially the bending wave speed, in order to obtain a converged solution, we select the ultrafine mesh (tetrahedral elements) to perform simulation calculations when using finite element software to visualize the vibration situation in the periodic plate.

The convergence solution is obtained by reducing the maximum element size. The remaining grid size parameters include: minimum element size 0.01mm, maximum element growth rate 1.3, curvature factor 0.2, and narrow area resolution 1. The convergence of the solution is introduced in FIG. S1. As the maximum unit size gradually decreases from 1.5mm, in the triangular lattice unit cell, the slope of the relationship curve between Dirac point frequency and maximum unit size (blue line) gradually decreases, which means that the calculation results gradually converge, and the corresponding convergence frequency value is about 2.06kHz. The green and orange lines respectively represent the degree of freedom of solution and the time taken to solve a single frequency point. It can be seen from the figure that when the maximum element size is less than 1 mm, the degree of freedom of solution and the time of solving a single frequency point increase rapidly. Considering that when the maximum unit size is 0.8mm, the Dirac point frequency is 2.066kHz, which is very close to the convergence solution of 2.06kHz, so the maximum element size is 0.8mm was used in the calculations described in the paper.

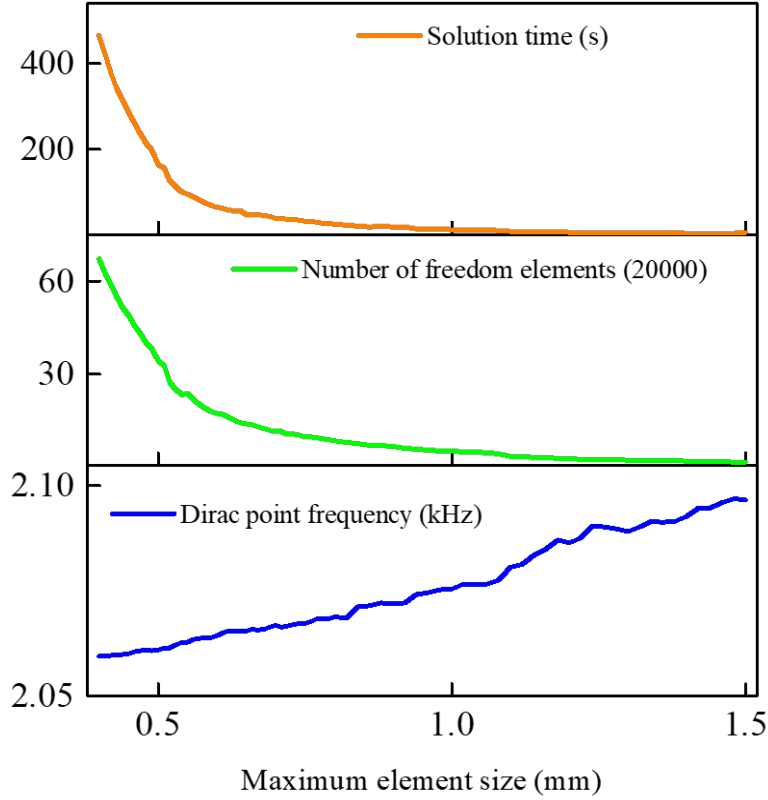

FIG. S1. The behavior of magnitudes of interest as a function of the maximum element size (in millimeters) employed in the finite element calculations.

## II. Local resonant modes induced by one-beam resonators

It has been demonstrated that in resonant cavities containing one-beam resonator, the rectangular beam acts like a spring[2]. Therefore, the changes in the band structure introduced by the one-beam resonators can be easily explained by using a simple spring-mass model, where the beam attaching the disk to the plate plays the role of the spring (see Fig. S2a). The model here developed is similar to that applied by Goffaux and Sanchez-Dehesa[3] to explain the dispersion relation caused by a lattice of local resonators in two-dimensional phononic materials. The model considers a lattice of masses connected by springs. Attached to each mass $M$, there exists a light pendulum (with mass $m$ and length $\ell$), which represents the localized mode associated to the one-beam resonator. This one-dimensional (1D) mechanical system has analytical solution and provides a physical insight of the changes in the dispersion relations for both in plane modes and out-of-plane modes produced by the hexagonal lattice of unit-cells described in Fig. S2a.

Figure S2b shows the dispersion relation of the out-of-plane modes (blue dotted lines) and the in-plane modes (red dotted lines) along the high symmetry directions in the Brillouin zone (BZ) of the hexagonal lattice. Let us first analyze the case of flexural (out-of-plane) modes. The band with an almost flat dispersion relation represents the band associated to the lattice of flexural local modes introduce by the one-beam resonator (see A in Fig. 3d in the manuscript). Note that a bandgap (grey region) occurs between the maximum frequency of the lower band, $f_b$ (located at the M point of the BZ), and the lowest frequency of the second band $f_u$ (located at the Γ point of the BZ). The bandwidth, $f_u - f_b$, gives an estimation of the interaction strength between the local resonant mode provided by the one-beam resonator and the continuum of flexural modes in the thin plate. Moreover, the interaction strength increases by increasing the mass of the one-beam resonator[3], also producing a decreasing of the mid bandgap frequency[2].

The same analysis can be extended to understand the dispersion relations represented as red dotted lines, where the flat band is associated to the lattice of translational in-plane modes, $f_{tr}$, excited in the one-beam resonators (see Fig. 3b in the manuscript). For these eigenmodes, the in-plane displacements of the one-beam resonators are much larger than that of the background plate. As for the case of flexural modes, a bandgap (grey region) appears because of the interaction between the local resonance and the in-plane waves propagating in the plate.

Regarding the other types of local resonances produced by the one-beam resonator, like the torsional modes ($f_{to}$), and rotational modes ($f_{ro}$), their dispersion relations can be calculated, but they are located at much higher frequencies[2], and they have no interest in the present work.

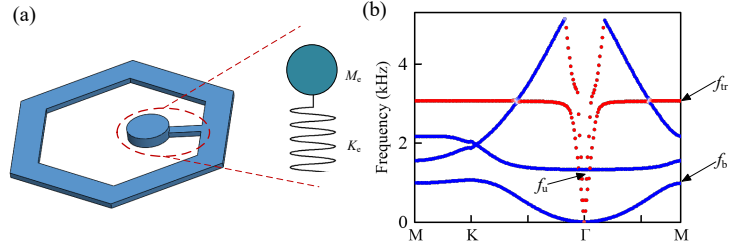

FIG. S2. (a) Schematic diagram of the unit cell containing a single one-beam resonator, enclosed by the red dashed circle. The resonator can be modeled by a spring mass system with effective mass $M_e$ and effective spring stiffness $K_e$. (b) The corresponding dispersion relation for the in-plane (red dotted lines) and out-of-plane (blue dotted lines) modes. The gray regions define the bandgap created by the corresponding local resonances.

## III. Edge states in zigzag ribbons

The intersection points between unit cells in the triangular lattice are abstracted as lattice points in a honeycomb lattice, as shown in Fig. S3(a), where $\boldsymbol{b_1} = (1/2, \sqrt{3}/2)a$ and $\boldsymbol{b_2} = (1/2, -\sqrt{3}/2)a$ represent the two basis vectors of the new honeycomb lattice. In addition, $\boldsymbol{\delta_i}$ ($i$ = 1, 2, 3) denotes the vector between two inequivalent lattice points in the honeycomb lattice. The reciprocal lattice vectors of the honeycomb lattice are denoted as $\boldsymbol{d_1} = 2\pi/a\,(1, 1/\sqrt{3})$, $\boldsymbol{d_2} = 2\pi/a\,(1, -1/\sqrt{3})$, as shown in Fig. S3(b). The Bloch wave vector $\boldsymbol{k}$ in the first Brillouin zone can be expressed as:

$$\boldsymbol{k} = k_h \boldsymbol{d_1} + k_v \boldsymbol{d_2},\ k_h, k_v \in [0,1). \tag{3}$$

Next, we investigate the effective Hamiltonian of graphene-like elastic plates $H(\boldsymbol{k})$ [4]:

$$H(\boldsymbol{k}) = [0, g(\boldsymbol{k});\ g^*(\boldsymbol{k}), 0]. \tag{4}$$

Considering the relative phase of adjacent lattice points in a honeycomb lattice, the off-diagonal term $g(\mathbf{k})$ can be expressed as[4]:

$$g(\mathbf{k}) = 1 + e^{i\mathbf{k}\cdot(\delta_3-\delta_2)} + e^{i\mathbf{k}\cdot(\delta_3-\delta_1)}. \tag{5}$$

Substituting Eq. (1) into Eq. (3),

$$g(k_h, k_v) = 1 + e^{-2\pi i k_h} + e^{2\pi i k_v}. \tag{6}$$

For an elastic graphene-like plate, $k_h$ can take arbitrary values ranging from 0 to 1. And the two-by-two effective Hamiltonian $H(\mathbf{k})$ can describe a one-dimensional lattice model $H(k_h)$, when $k_v$ runs through the one-dimensional Brillouin zone for each $k_h$. Therefore, the winding number can be defined as[5]:

$$e(k_h) = \frac{1}{2\pi i}\int_0^1 \frac{\partial \ln[g(k_h, k_v)]}{\partial k_v} dk_v, \tag{7}$$

where $g(k_h, k_v)$ traces a closed curve in the complex plane, and $e(k_h)$ represents the number of times it winds around the origin. Additionally, chiral symmetry also ensures the quantized Zak phase[6]:

$$\theta(k_h) = \int_0^1 i\langle\psi(k_v)|\nabla_{k_v}\psi(k_v)\rangle dk_v, \tag{8}$$

which relates to the winding number through the relation[5]:

$$\theta(k_h) = |e(k_h)|\pi. \tag{9}$$

The theoretically calculated Zak phase, as shown by the blue line (Theory) in Fig. 5c, reveals that in the interval of $1/3 < k_h < 2/3$, the Zak phase of the elastic plate with zigzag boundaries is π, which implies that the system is topologically non-trivial. In the simulation calculations, normalized displacements are used as eigenstates $\langle\psi(\mathbf{k})|$, and the Wilson-loop method is employed to compute the Zak phase of the system. Discretize $\mathbf{d}_1$ and $\mathbf{d}_2$ into $N_1$ and $N_2$ equal segments, and satisfy the relation:

$$\begin{cases} k_{h,n} = \frac{n}{N_1} \\ k_{v,m} = \frac{m}{N_2} \end{cases}, \quad n \in [0, 1, \quad N_1\text{-}1),\, m \in [0, 1, \quad N_2\text{-}1). \quad (10)$$

For each $k_{h,n}$, the Zak phase is the sum of the Berry phases in all small segments (from $k_{v,m}$ to $k_{v,m+1}$) in the $d_2$ direction, which can be expressed as:

$$e^{-i\theta_{k_h}} = \prod_{m=0}^{N_2-1} e^{-i\theta_{k_{v,m}}} = \prod_{m=0}^{N_2-1} \langle \psi(k_{v,m}) | \psi(k_{v,m+1}) \rangle. \quad (11)$$

By calculating the Zak phase of the Bloch wave function represented by displacement on the compressed one-dimensional space, as shown by the red dot (FEM) in Fig. 5c, the distribution of Zak phase on the fourth band (located near the Dirac point) is same with the theoretical calculated value, i.e., in the interval of $1/3 < k_h < 2/3$, the system has a non-trivial phase and supports topological edge states. In fact, within the interval of $k_h$, the transition point between topological trivial phase and topological non-trivial phase corresponds to the corner points K and K' in the first Brillouin zone. Therefore, in the projected band structure, the edge states located between K and K' are topologically protected.

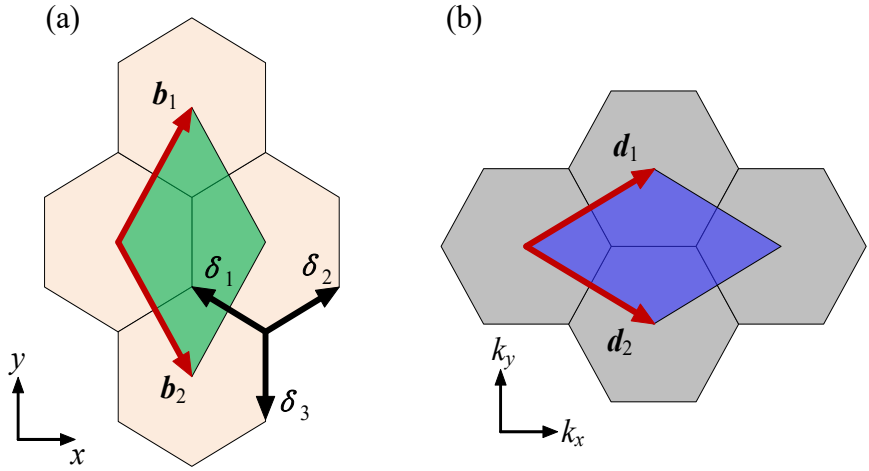

FIG. S3 (a) The unit cell of the equivalent honeycomb lattice in the super unit cell of the triangular lattice. The orange color in the figure represents the triangular lattice, and the rhombus green area represents the unit cell of the honeycomb lattice with $b_1$ and $b_2$ as basis vectors; the vectors connecting the adjacent non-equivalent lattice points in the honeycomb lattice are represented by $\delta_i$ ($i=1, 2, 3$). (b) The reciprocal lattice and the first Brillouin zone of the honeycomb lattice. The blue shading represents the first Brillouin zone with $d_1$ and $d_2$ as basis vectors.

## IV. Edge states in zigzag ribbons

A double degenerate band occurs at the frequency of the Dirac point, and the edge states on this band are topologically protected. Figure S4 shows numerical simulations demonstrating that the edge states contained in the band structure of ribbons are independent of the number N of Cells III contained in the 1D ribbon.

The left panels in Fig. S4 shows the band structure obtained with supercell composed of two (S4a), three (S4b), four (S4c) and five (S4d) Cell III, respectively. The right panels represent the corresponding out-of-plane displacement fields of the eigenmodes with frequencies belonging to the up and down bands containing edge states at $k_x$=0.9. the displacement in Fig. S4 show that the out-of-plane displacements are always localized at the top or bottom boundaries of the ribbon, in spite of the fact that this ribbon is composed of just two cells type III. The fact that ribbons with smaller dimensions are capable to generate edge states provides a feasible solution for greatly reducing the sample size and realizing functional integration design and other engineering applications.

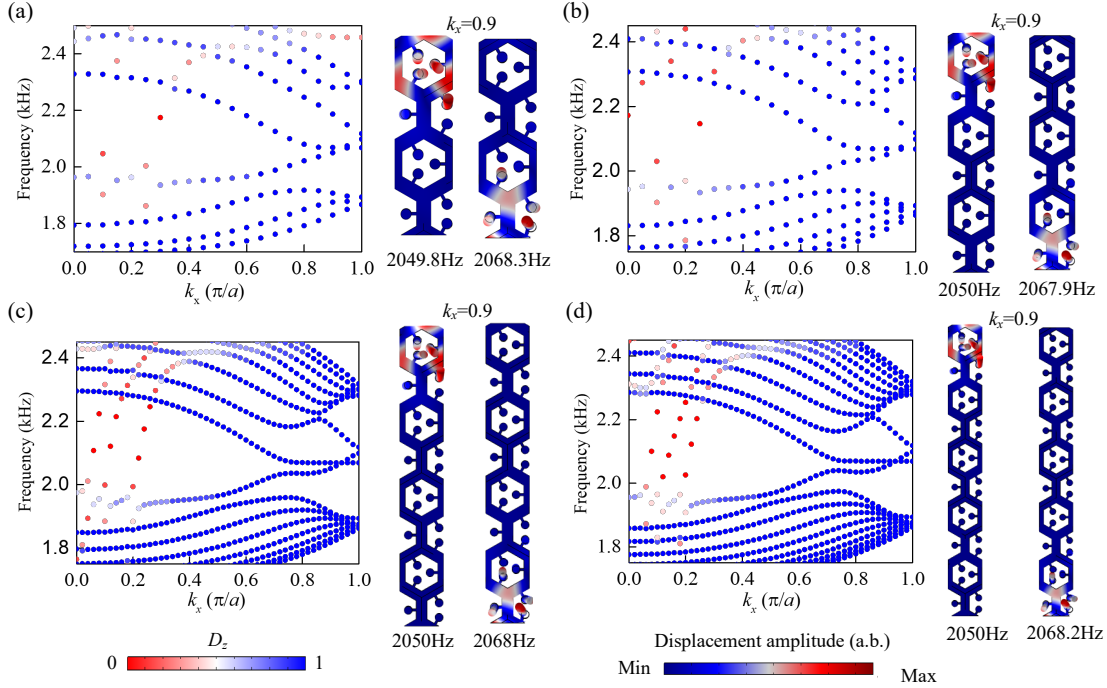

FIG. S4 (a)-(e) Band structure and out-of-plane displacement fields of edge states at $k_x=0.9$ calculated for ribbons composed of two, three, four, five and six Cell III, respectively. The color in the band structure represents the out-of-plane mode polarizability, and the color in the modal distribution represents the displacement amplitude distribution.

## V. Propagation in metamaterial plate

To analyze the propagation of flexural waves in a metamaterial plate, a thin plate consisting of 12 supercells containing 5 Cell II was selected as shown in Fig. S5. A low reflection boundary condition is applied to the left and right boundary of the metamaterial plate, and the upper and lower boundary remain free.

We apply harmonic vibration excitation at the upper part of the plate, represented by the green circles in the illustration. Figure S5 displays the distribution of out-of-plane velocity fields for excitation source frequencies of 2020Hz, 2040Hz, 2050Hz, 2060Hz, 2070Hz, 2080Hz, 2090Hz, 2100Hz and 2200Hz. From the figure, it can be observed that when the excitation source frequency is below 2050Hz or above 2090Hz, the vibration can extend throughout the entire metamaterial plate. When the frequency is between 2050Hz and 2090Hz (with a bandwidth of approximately 40Hz), flexural wave vibrations propagate along the free boundary of the metamaterial plate, demonstrating robustness in transmission. In addition, the vibration excited by the point source with frequency of 2070Hz can not only propagate along the top boundary of the plate, at the bottom boundary of the plate, there is a local mode vibration. This is because the eigenmodes (near the Dirac point) at frequency of 2070 Hz exhibits a strong local resonance pattern at the bottom end of the ribbon, as shown in Fig. 4e. Therefore, the weak vibration propagating to the bottom boundary is strengthened. In fact, when the frequency is near the Dirac point, such as 2070Hz, once the excitation is applied to the top end of the metamaterial plate, there will also be weak local vibration in the bottom end of the metamaterial plate. This phenomenon can be used to judge the Dirac point frequency of the system in practical applications, and is useful to estimate the range of edge states.

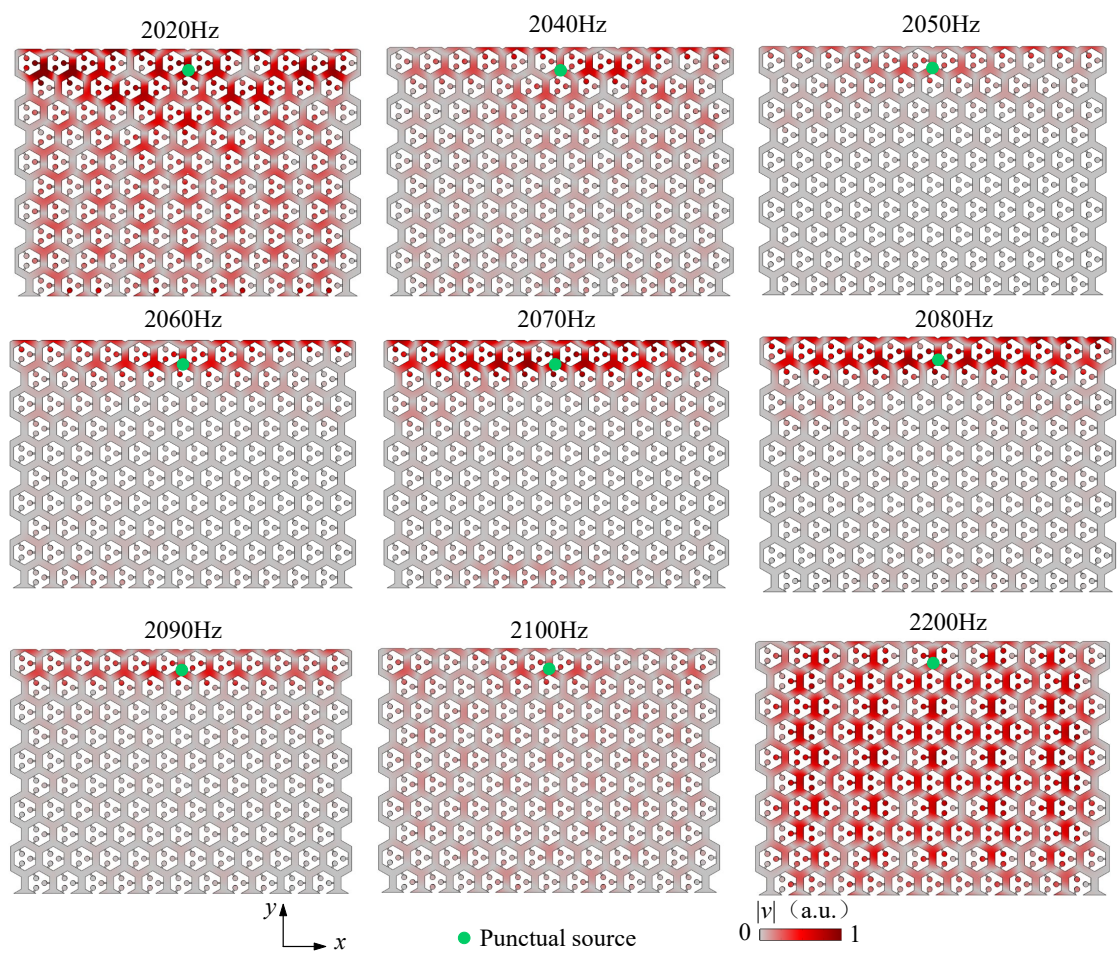

FIG. S5. The distribution of out-of-plane velocity fields for different excitation source frequencies. The green pentagram represents the point source.